\documentclass[12pt,english,floatfix,superscriptaddress,aps,prd,preprint,showkeys,nofootinbib]{revtex4}
\usepackage{amsmath}
\usepackage{amssymb}
\usepackage{amsbsy}
\usepackage{amsfonts}
\usepackage{amsopn}
\usepackage{amstext}
\usepackage{graphicx}
\usepackage[english]{babel}
\usepackage{color}
\usepackage{slashed}
\usepackage{esint}
\usepackage[dvips]{epsfig}
\usepackage[dvips]{graphicx}
\usepackage{float}
\usepackage{units}
\usepackage{textcomp}
\usepackage{wasysym}
\usepackage{hyperref}
\usepackage{slashed}

\begin{document}

\title{Duality between the Maxwell-Chern-Simons and self-dual models in very special relativity}

\author{Fernando M. Belchior}
\email{belchior@fisica.ufc.br}
\affiliation{Universidade Federal do Cear\'a (UFC), Departamento de F\'isica,\\ Campus do Pici, Fortaleza - CE, C.P. 6030, 60455-760 - Brazil.}

\author{Roberto V. Maluf}
\email{r.v.maluf@fisica.ufc.br}
\affiliation{Universidade Federal do Cear\'a (UFC), Departamento de F\'isica,\\ Campus do Pici, Fortaleza - CE, C.P. 6030, 60455-760 - Brazil.}
\affiliation{Departamento de F\'{i}sica Te\'{o}rica and IFIC, Centro Mixto Universidad de Valencia - CSIC. Universidad
de Valencia, Burjassot-46100, Valencia, Spain.}

\begin{abstract}
This work aims to investigate the classical-level duality between the $SIM(1)$-Maxwell-Chern-Simons (MCS) model and its self-dual counterpart. Initially, our focus is on free-field cases to establish equivalence through two distinct approaches: comparing the equations of motion and utilizing the master Lagrangian method. In both instances, the classical correspondence between the self-dual field and the MCS dual field undergoes modifications due to very special relativity (VSR). Specifically, duality is established only when the associated VSR-mass parameters are the same. Furthermore, we analyze the duality when the self-dual model is minimally coupled to fermions. As a result,  we show that Thirring-like interactions, corrected for non-local VSR contributions, are included in the MCS model. Additionally, we demonstrate the equivalence of the fermion sectors in both models, thereby concluding the proof of classical-level duality.
\end{abstract}

\maketitle

%%%%%%%%%%%%%%%%%%%%%%%%%%%%%%%%%%%%%%%%%%%%%%%%%%%%%%%%%%%%%%%%%%%%%%%%%%%%%%%%%%%%%%%%%%%
\section{Introduction}

The physics beyond the Standard Model has been increasingly investigated in recent years. Despite being an important and notable symmetry of nature, Lorentz symmetry might experience small deviations at the Planck scale, as suggested by string theory \cite{Kostelecky:1988zi}. To understand such deviations, the Standard Model extension (SME) \cite{Carroll:1989vb, Colladay:1996iz, Colladay:1998fq} emerges as a powerful theoretical framework. It was conceived as an effective field theory containing Lorentz-violating terms.

The possibility of Lorentz symmetry violation has been speculated in theories beyond string theory, such as Horava-Lifshitz \cite{Horava:2009uw}, noncommutatives \cite{Carroll:2001ws}, and very special relativity (VSR) \cite{Cohen:2006ky}. In the latter, the complete Lorentz group is broken down into subgroups that still preserve conservative laws, as well as the effects predicted by special relativity. These subgroups are the Homothety group $HOM(2)$ and the similitude group $SIM(2)$ \cite{Lee:2015tcc}. Additionally, there is a preferred direction in space-time due to the presence of the null vector $n_{\mu}$. Another important ingredient of VSR is the presence of nonlocal operators that overcome potential issues related to the violation of unitarity and causality. As a consequence, it is possible to observe mass generation for fermionic fields \cite{Fan:2006nd} and also for gauge fields, such as the photon \cite{Alfaro:2013uva, Alfaro:2019koq}. In this case, there would be a massive gauge field with two degrees of freedom, which is gauge-invariant, unlike the Proca theory \cite{Cheon:2009zx}. VSR also opens the theoretical possibility for explaining neutrino mass \cite{Cohen:2006ir} and dark matter \cite{Ahluwalia:2010zn}.

Investigations into dualities and equivalence between seemingly different theories have recently garnered considerable attention in both high-energy and condensed-matter physics \cite{Mandelstam:1975hb, Witten:1983ar, Marino:1990yi}. An intriguing example in this context is the $AdS/CFT$ correspondence, proposed by Juan Maldacena, which connects a string theory in 10 dimensions to conformal field theory (CFT) \cite{Maldacena:1997re}. The $T$ and $S$ dualities, which also appear in string theory \cite{Strominger:1996it, Polchinski:2014mva}, contribute to the richness of this theory. Furthermore, the topic of duality also appears as an important technique to study non-perturbative aspects of field theories in low dimensions \cite{Burgess:1993np}. More recently, the application of bosonization has given rise to a novel particle-vortex duality in $2+1$ dimensions known as the duality web \cite{Nastase:2017gxr, Hernaski:2018jsy}.

In the context of topologically massive theories, Deser and Jackiw were the first to propose an equivalence between the Maxwell-Chern-Simons (MCS) and self-dual (SD) models in $2+1$ dimensions \cite{Deser:1984kw}. Following that seminal work, several subsequent studies have been published, exploring issues such as coupling with matter \cite{Ilha:2001he, Anacleto:2000ea, Dalmazi:2006yv} and supersymmetric extensions \cite{Ferrari:2008he, Ferrari:2006vy}. Notably, when considering the self-dual model minimally coupled to Dirac fermions, a Thirring term must be added to the MCS Lagrangian to preserve the equivalence \cite{Gomes:1997mf}. This equivalence can be demonstrated through two different methods: gauge embedding \cite{Anacleto:2001rp} and the master action \cite{Dalmazi:2008zh}. 

In the first approach, gauge symmetry is revealed through a process of embedding in the SD model, leading to MCS \cite{Marques:2022hhl}. On the other hand, the master action involves constructing a first-order derivative Lagrangian to interpolate between the two theories. This method has the advantage of preserving gauge invariance and facility the proof of duality at the quantum level. This topic has also been discussed in the context of Lorentz symmetry violation \cite{BaetaScarpelli:2014jfl, Furtado:2008gs, Guimaraes:2010cu, Fargnoli:2014lca}. Additionally, through the $B\wedge F$ term, it is possible to extend this duality to $3+1$ dimensions \cite{Menezes:2002be, Maluf:2020fch, Gama:2022nue}. Such a topologically massive theory is constructed by considering a Lagrangian density composed of a vector field $A_\mu$ and an antisymmetric 2-tensor field $B_{\mu\nu}$ known as the Kalb-Ramond field \cite{Kalb:1974yc, Maluf:2023gbf}.

In the three-dimensional spacetime, the $SIM(1)$ version of the MCS theory was proposed by Bufalo in Ref. \cite{Bufalo:2016lfq}. Afterward, the VSR contributions to the induced Maxwell-Chern-Simons terms for the effective action at one-loop order were carried out in Ref. \cite{Bufalo:2020cst}. Also, the construction of the effective action in the non-Abelian case was addressed in Ref. \cite{Haghgouyan:2022dvq}. Here, we are interested in investigating the duality between the Maxwell-Chern-Simons and the self-dual models in the framework of very special relativity. Firstly, the VSR-based MCS and SD models are discussed, demonstrating equivalence at the level of equations of motion in the free-field case. Additionally, through the master Lagrangian approach, we provide direct proof of equivalence at the classical level. Moreover, we include minimal coupling with fermionic matter and analyze whether the duality is maintained.

This paper is structured as follows: In Sec. (\ref{s1}), we aim to present the theoretical models, computing the equations of motion and the associated Feynman propagators. Subsequently, we establish equivalence at the level of equations of motion by considering only the free-field case. In Sec. (\ref{s2}), we construct the master Lagrangian by introducing an auxiliary field. In Sec. (\ref{s3}), we introduce the fermion Lagrangian in the framework of VSR and then couple the self-dual model to fermions. By employing the master Lagrangian, we demonstrate that a Thirring-like term, involving VSR non-local corrections, appears in the MCS model. We show classical equivalence in both the gauge and matter sectors.

%%%%%%%%%%%%%%%%%%%%%%%%%%%%%%%%%%%%%%%%%%%%%%%%%%%%%%%%%%%%%%%%%%%%%%%%%%%%%%%%%%%%%%%%%%
\section{Maxwell-Chern-Simons and self-dual models in VSR}\label{s1}

We started our analysis by defining the topologically $SIM(1)$–invariant Maxwell–Chern–Simons model described by \cite{Bufalo:2016lfq}:
\begin{align}\label{l1}
\mathcal{L}_{MCS}=-\frac{1}{4}\widetilde{F}_{\mu\nu}\widetilde{F}^{\mu\nu} +\frac{1}{4}\theta\epsilon^{\mu\nu\rho}A_\mu\widetilde{F}_{\nu\rho}.
\end{align}Here, $\widetilde{F}_{\mu\nu}$ represents the field strength tensor associated with the gauge field $A_{\mu}$ and is defined as
\begin{equation}
   \widetilde{F}_{\mu\nu}=\widetilde{\partial}_\mu A_\nu-\widetilde{\partial}_\nu A_\mu, 
\end{equation}where the $SIM(1)$ wiggle operator is given by
\begin{align}
\widetilde{\partial}_\mu=\partial_\mu+\frac{m_{A}^2}{2}\frac{ n_\mu}{(n\cdot \partial)},
\end{align}and $m_{A}$ represents the VSR-mass related to the $A_\mu$ field. The constant null vector $n_\mu=(1,0,1)$ defines a preferred direction in the spacetime and transforms multiplicatively covariant under the $SIM(1)$ group \cite{Vohanka:2014lsa}.

The gauge symmetry of the VSR-MCS model is expressed as  
\begin{equation}
    A_{\mu}\rightarrow A_{\mu}+\widetilde{\partial}_\mu\Lambda,
\end{equation}where $\Lambda(x)$ represents an arbitrary gauge parameter. The field strength $\widetilde{F}_{\mu\nu}$ can be written as
\begin{align}\label{tfs}
\widetilde{F}_{\mu\nu}=\mathcal{F}_{\mu\nu}+\frac{1}{2}m_{A}^2\Big[\frac{n_\mu}{(n\cdot \partial)^2}n^\lambda\mathcal{F}_{\lambda\nu}-\frac{n_\nu}{(n\cdot \partial)^2}n^\lambda\mathcal{F}_{\lambda\mu}\Big],
\end{align}
where we introduce the auxiliary tensor \cite{Maluf:2023gbf}
\begin{align}
\mathcal{F}_{\mu\nu}=F_{\mu\nu}+\frac{1}{2}m_{A}^2\Big[\frac{n_\mu}{(n\cdot \partial)^2}n^\lambda\partial_\nu A_\lambda-\frac{n_\nu}{(n\cdot \partial)^2}n^\lambda\partial_\mu A_\lambda\Big],
\end{align}
with $F_{\mu\nu}=\partial_\mu A_\nu-\partial_\nu A_\mu$ being the standard field strength. It is important to note that the tensor $\mathcal{F}_{\mu\nu}$ reduces to standard field strength $F_{\mu\nu}$ when we make the following shift
\begin{align}
A_\mu\rightarrow A_\mu+\frac{m_{A}^2}{2}\frac{ n_\mu}{(n\cdot \partial)^2} (n\cdot A).  
\end{align}

We can utilize the wiggle field strength (\ref{tfs}) along with the above field redefinition to express the VSR-MCS Lagrangian as
\begin{align}\label{lm}
\mathcal{L}_{MCS}=-\frac{1}{4} F_{\mu\nu}F^{\mu\nu}+\frac{1}{4}\theta\epsilon^{\mu\nu\rho}A_\mu F_{\nu\rho}+\frac{1}{2}m_{A}^2 n_\mu\Big(\frac{1}{n\cdot\partial}F^{\mu\nu}\Big)n^\lambda\Big(\frac{1}{n\cdot\partial}F_{\lambda\nu}\Big)\nonumber\\+\frac{1}{8}\theta m_{A}^2\epsilon^{\mu\nu\rho}F_{\nu\rho}\frac{n_\mu}{(n\cdot \partial)^2}\partial_\rho (n\cdot A) +\frac{1}{4}\theta m_{A}^2\epsilon^{\mu\nu\rho}A_\mu\Big[\frac{n_\nu}{(n\cdot \partial)^2}n^\lambda F_{\lambda\rho}\Big].
\end{align}
As we can verify, the VSR-MCS action remains invariant under the standard gauge transformation $\delta A_\mu = \partial_\mu\Lambda$, even in the presence of additional terms containing the massive parameter $m_{A}$.

The equation of motion for $A_\mu$ is obtained from the Lagrangian density (\ref{lm}). Explicitly, we find 
\begin{align}
\partial_\mu F^{\mu\nu}+m_A ^2 \frac{n_\mu}{(n\cdot\partial)}F^{\mu\nu} -\frac{1}{2}m_A ^2 \frac{n^\nu}{(n\cdot\partial)^2}\partial_\mu(n_\alpha F^{\alpha\mu}) + \frac{1}{2} \theta\epsilon^{\nu\mu\rho}F_{\mu\rho}\nonumber\\+\frac{1}{2}\theta m_{A}^2\epsilon^{\nu\mu\rho}\Big[\frac{n_\mu}{(n\cdot \partial)^2}n^\alpha F_{\alpha\rho}\Big]=0.\label{EoMMCS}
\end{align}

For future analysis, it is convenient to rewrite the equation of motion (\ref{EoMMCS}) in terms of the dual vector associated with the field strength $F_{\mu\nu}$, defined by
\begin{equation}
    F^\mu=\frac{1}{2\theta}\epsilon^{\mu\nu\rho}F_{\nu\rho}.\label{dualvector}
\end{equation}Then, the EoM (\ref{EoMMCS}) assumes the following expression:
\begin{align}\label{Fe}
\theta\Big[ F^\mu-\frac{m_{A}^2}{2}\frac{n^\mu n_\lambda}{(n\cdot\partial)^2}F^\lambda\Big]-\epsilon^{\mu\nu\rho}\partial_\nu F_\rho -\frac{1}{2}m_{A}^2\epsilon^{\mu\nu\rho}\frac{n_\rho n_\lambda}{(n\cdot\partial)^2}\partial_\nu F^\lambda+\frac{1}{2}m_{A}^2\epsilon^{\mu\nu\rho}\frac{n_\nu}{(n\cdot\partial)}F_\rho =0.
\end{align}

Continuing with our analysis of the MCS model, we aim to find the Feynman propagator in order to determine the spectrum of the theory. To achieve this, a gauge fixing is required, and one proceeds by writing:
\begin{align}
    \mathcal{L}=\mathcal{L}_{MCS}-\frac{1}{2}\bigg(\partial_\mu A^\mu+\frac{m_{A}^2}{2}\frac{n_\mu}{(n\cdot \partial)}A^\mu\bigg).
\end{align}
Thus, we can express the above Lagrangian density in the bilinear form as follows
\begin{align}
\mathcal{L}=\frac{1}{2}A_\mu\mathcal{O}^{\mu\nu}A_\nu,   
\end{align}where $\mathcal{O}^{\mu\nu}$ stands for the associated wave operator given in the momentum space ($k_\mu=i\partial_\mu$) by
\begin{align}
\mathcal{O}^{\mu\nu}= -(k^2-m_{A}^2)\theta^{\mu\nu}-(k^2-m_{A}^2)\omega^{\mu\nu} -\theta S^{\mu\nu}+\frac{\theta m_{A}^2}{2}\bigg[\frac{L^{\mu\nu}}{(n\cdot k)}-\frac{Q^{\mu\nu}-Q^{\nu\mu}}{(n\cdot k)^2}\bigg].
\end{align}Here, we have introduced the set of spin-projection operators defined by
\begin{align}
 \theta^{\mu\nu}=\eta^{\mu\nu}-\omega^{\mu\nu}\ ,\ \omega^{\mu\nu}=\frac{k^\mu k^\nu}{k^2}\ ,\ S^{\mu\nu}=i\epsilon^{\mu\nu\lambda}k_\lambda,\\
 L^{\mu\nu}=i\epsilon^{\mu\nu\lambda}n_\lambda\ ,\
 L^{\mu\nu}=n^\mu T^\nu\ ,\
 \Sigma^{\mu\nu}= n^\mu k^\nu,\\
 \Lambda^{\mu\nu}= n^\mu n^\nu\ ,\
 \Phi^{\mu\nu}=T^\mu k^\nu \ ,\
 M^{\mu\nu}=T^\mu T^\nu,\\
T^\mu=i\epsilon^{\mu\nu\lambda}k_\nu n_\lambda,
\end{align}such that these operators satisfy a closed algebra when their products are considered.

The Feynman propagator $\Delta_{\rho\mu}(k)$ is obtained by inverting the equation
\begin{align}
\mathcal{O}^{\mu\nu}(k)\Delta_{\rho\mu}(k)=i\delta^\nu_\rho.   
\end{align}

Thereby, after some algebraic manipulations, we arrive at the following expression for the VSR Maxwell-Chern-Simons field propagator:
\begin{align}
\Delta_{\rho\mu}(k)=\frac{-i}{k^2-m_{A}^2-\theta^2}\bigg(\eta_{\rho\mu}-\frac{k_\rho k_\mu}{(k^2-m_{A}^2)}+\frac{m_{A}^2}{2}\frac{n_\rho k_\mu+n_\mu k_\rho}{(n\cdot k)(k^2-m_{A}^2)}-\frac{m_{A}^4}{4}\frac{n_\rho n_\mu}{(n\cdot k)^2(k^2-m_{A}^2)}\nonumber\\-i\theta\epsilon_{\rho\mu\lambda}\frac{k^\lambda}{(k^2-m_{A}^2)}+\frac{\theta m_{A}^2}{2}\epsilon_{\rho\mu\lambda}\frac{n^\lambda}{(n\cdot k)(k^2-m_{A}^2)}\bigg)-\frac{k_\rho k_\mu}{(k^2-m_{A}^2)^2}\nonumber\\+\frac{m_{A}^2}{2}\frac{n_\rho k_\mu+n_\mu k_\rho}{(n\cdot k)(k^2-m_{A}^2)^2}-\frac{m_{A}^4}{4}\frac{n_\rho n_\mu}{(n\cdot k)^2(k^2-m_{A}^2)^2}.  \label{MCSprop}
\end{align}

From the above expression, we can identify the gauge field mass through the pole at $k^2-m_{A}^2-\theta^2=0$. Additionally, there is a pole at $(n\cdot k)=0$, indicating a non-physical mode. To handle these types of poles, we can employ the Alfaro-Mandelstam-Leibbrandt prescription \cite{Alfaro:2016pjw}, which addresses UV/IR mixing divergences in one-loop Feynman integrals. Furthermore, setting $m_{A}=0$ in Eq. (\ref{MCSprop}) allows us to recover the standard MCS propagator \cite{Gomes:1997mf}.

Once we have introduced the Maxwell-Chern-Simons theory in VSR, we are interested now in studying its connection with the VSR-inspired self-dual (SD) model. Thus, we propose the following first-order derivative Lagrange density 
\begin{align}\label{sd}
\mathcal{L}_{SD}=\frac{1}{2}\theta^2 f_\mu f^\mu-\frac{1}{4}\theta\epsilon^{\mu\nu\lambda}f_\mu\widehat{f}_{\nu\lambda},
\end{align}
where we have defined the associated field strength by  $\widehat{f}_{\mu\nu}=\widehat{\partial}_\mu A_\nu-\widehat{\partial}_\nu A_\mu$, and the hat derivative operator as
\begin{align}
\widehat{\partial}_\mu=\partial_\mu+\frac{m_{f}^2}{2}\frac{ n_\mu}{(n\cdot \partial)}.\label{hatderivative}
\end{align} 

Note that $m_{f}$ is the VSR-mass associated to the self-dual field $f_\mu$. From the definition (\ref{hatderivative}), we can write the VSR-self-dual Lagrangian (\ref{sd}) as 
\begin{align}\label{sd2}
\mathcal{L}_{SD}=\frac{1}{2}\theta^2 f_\mu f^\mu-\frac{1}{2}\theta\epsilon^{\mu\nu\lambda}f_\mu\partial_\nu f_\lambda--\frac{1}{4}\theta m_{f}^2\epsilon^{\mu\nu\lambda}f_\mu\frac{n_\nu}{(n\cdot \partial)} f_\lambda.
\end{align}

In contrast to the gauge-invariant VSR-MCS model (\ref{lm}), the Lagrangian density for VSR-SD (\ref{sd2}) lacks gauge invariance. The equation of motion for the SD field $f_\mu$ is readily obtained from (\ref{sd2}), 
\begin{align}\label{fe}
\theta f^\mu - \epsilon^{\mu\nu\rho}\partial_{\nu}f_\rho-\frac{1}{2} m_{f}^2\epsilon^{\mu\nu\rho}\frac{n_\nu}{(n\cdot\partial)}f_\rho=0.
\end{align}

A direct comparison between equations (\ref{Fe}) and (\ref{fe}) reveals that the dual field $f_\mu$ follows the exact equation of motion obtained for the VSR-MCS model when we set $m_{f}=m_{A}$ and make the identification
\begin{align}\label{fF}
f^\mu\leftrightarrow F^\mu-\frac{m_{A}^2}{2}\frac{n^\mu n_\lambda}{(n\cdot\partial)^2}F^\lambda.   
\end{align}

Hence, the fundamental field in the VSR-SD model is identified as the dual field (\ref{dualvector}) of the MCS model, added by a nonlocal VSR correction. This establishes classical equivalence through the equations of motion for the free field case. It is important to note that the result obtained by Deser and Jackiw is recovered when we cancel out the VSR mass parameter.

Finally, we can analyze the spectrum of VSR-SD theory by computing the associated Feynman propagator. By employing the same algebra presented previously, the propagator for the self-dual field results in
\begin{align}
 D_{\rho\mu}(k)= \frac{-i}{k^2-m_{f}^2-\theta^2}\bigg(\eta_{\rho\mu}-\frac{k_\rho k_\mu}{\theta^2}+\frac{m_{f}^2}{2\theta^2}\frac{n_\rho k_\mu+n_\mu k_\rho}{(n\cdot k)}-\frac{m_{f}^4}{4\theta^2}\frac{n_\rho n_\mu}{(n\cdot k)^2}\nonumber\\-i\epsilon_{\rho\mu\lambda}\frac{k^\lambda}{\theta}+\frac{m_{f}^2}{2\theta}\epsilon_{\rho\mu\lambda}\frac{n^\lambda}{(n\cdot k)}\bigg).  
\end{align}
It is worth noting the presence of the massive pole at $k^2-m_{f}^2-\theta^2=0$ and the non-physical pole at $(n\cdot k)=0$, similar to the VSR-MCS propagator. As in the previous case, if we set $m_{f}=0$, we recover the Lorentz covariant result \cite{Gomes:1997mf}.

%%%%%%%%%%%%%%%%%%%%%%%%%%%%%%%%%%%%%%%%%%%%%%%%%%%%%%%%%%%%%%%%%%%%%%%%%%%%%%%%%%%%%%%%%%%%
\section{Master lagrangian approach}\label{s2}

In the previous section, we established a connection between MCS and SD models within the framework of very special relativity at the level of equations of motion. Now, employing the master Lagrangian method enables the interpolation between these models. This approach provides a more direct verification of duality at the quantum level. The first step involves introducing an auxiliary field, transforming the Lagrangian density $\mathcal{L}_{MCS}$ into a first-order derivative form \cite{Maluf:2020fch}:
\begin{align}\label{mlf}
    \mathcal{L}_{M}=a\epsilon^{\mu\nu\rho}\Pi_\mu\partial_\nu A_\rho+\frac{a m_{A}^2}{2}\epsilon^{\mu\nu\rho}\Pi_\mu\frac{n_\rho}{(n\cdot\partial)^2}\partial_\nu (n\cdot A) + \frac{a m_{A}^2}{2}\epsilon^{\mu\nu\rho}\Pi_\mu \frac{n_\nu}{(n\cdot \partial)}A_\rho+b\Pi_\mu \Pi^\mu\nonumber\\+\frac{\theta}{2}\epsilon^{\mu\nu\rho}A_\mu\partial_\nu A_\rho-\frac{\theta m_{A}^2}{2}\epsilon^{\mu\nu\rho}A_\mu \frac{n_\nu}{(n\cdot \partial)^2}\partial_\rho(n\cdot A)+\frac{\theta m_{A}^2}{4}\epsilon^{\mu\nu\rho}A_\mu\frac{n_\nu}{(n\cdot \partial)}A_\rho,
\end{align}
where $a$ and $b$ are constant coefficients to be determined. As we can observe, the master Lagrangian density is written in terms of the fields $A_{\mu}$ and $\Pi_{\mu}$. The absence of the derivative term and the presence of the mass term for $\Pi_{\mu}$ guarantees the auxiliary nature of this field.

By varying the action associated with $\mathcal{L}_{M}$, i.e., $\int d^3x \mathcal{L}_{M}$, with respect to $\Pi_\mu$, we obtain the equations of motion for the auxiliary field:
\begin{equation}
\Pi^\mu=-\frac{a}{2b}\bigg(\epsilon^{\mu\nu\rho}\partial_\nu A_\rho+\frac{m_{A}^2}{2}\epsilon^{\mu\nu\rho}\frac{n_\rho}{(n\cdot \partial)^2}\partial_\nu (n\cdot A)+\frac{m_{A}^2}{2}\epsilon^{\mu\nu\rho}\frac{n_\nu}{(n\cdot \partial)}A_\rho\bigg ). \label{ppf}    
\end{equation}

A similar procedure can be done for the field $A_{\mu}$, and we can solve their equations of motion, leading the following solution:
\begin{equation}
   A_\mu+\frac{m_{A}^2}{2}\frac{n_\mu}{(n\cdot\partial)^2}(n\cdot A)=-\frac{a}{\theta}\Pi_\mu+\partial_\mu \Sigma +\frac{n_\mu}{(n\cdot \partial)} \Sigma,\label{af}  
\end{equation}where $\Sigma$ is an arbitrary scalar field.

It is easy to verify that substituting (\ref{ppf}) into (\ref{mlf}) and imposing $\mathcal{L}_{M}=\mathcal{L}_{MCS}$, we find the relations
\begin{align}
a=\pm\theta,\\
b=\frac{\theta^2}{2},
\end{align}and the sign $\chi=\pm1$ determines either the self-duality $(+)$ or anti-self-duality $(-)$ of the theory. Now, replacing (\ref{af}) in (\ref{mlf}), we recover the VSR-SD Lagrangian $\mathcal{L}_{SD}$. 

Now, we can fix $a=\chi\theta$ and $b=\theta^2/2$ in (\ref{mlf}) so that our master Lagrangian takes the final form
\begin{align}\label{mlf1}
\mathcal{L}_{M}=\chi\theta\epsilon^{\mu\nu\rho}\Pi_\mu\partial_\nu A_\rho+\frac{\chi\theta m_{A}^2}{2}\epsilon^{\mu\nu\rho}\Pi_\mu\frac{n_\rho}{(n\cdot\partial)^2}\partial_\nu (n\cdot A) + \frac{\chi\theta m_{A}^2}{2}\epsilon^{\mu\nu\rho}\Pi_\mu \frac{n_\nu}{(n\cdot \partial)}A_\rho+\frac{\theta^2}{2}\Pi_\mu \Pi^\mu\nonumber\\+\frac{\theta}{2}\epsilon^{\mu\nu\rho}A_\mu\partial_\nu A_\rho-\frac{\theta m_{A}^2}{2}\epsilon^{\mu\nu\rho}A_\mu \frac{n_\nu}{(n\cdot \partial)^2}\partial_\rho(n\cdot A)+\frac{\theta m_{A}^2}{4}\epsilon^{\mu\nu\rho}A_\mu\frac{n_\nu}{(n\cdot \partial)}A_\rho. 
\end{align}
On the other hand, the equation of motion (\ref{ppf}) can be rewritten using $F^\mu=-\frac{1}{\chi\theta}\epsilon^{\mu\nu\rho}\partial_\nu A_\rho$ as 
\begin{align}
\Pi^\mu=F^\mu-\frac{m_{A}^2}{2}\frac{n^\mu n_\lambda}{(n\cdot\partial)^2}F^\lambda,    
\end{align}which is the same relation obtained in (\ref{fF}).

As mentioned earlier, the master Lagrangian (\ref{mlf1}) gives rise to both the MCS and SD models in the VSR scenario. Moreover, it is easily verified that $\mathcal{L}_{M}$ is gauge-invariant under $\delta A_\mu=\tilde{\partial}_{\mu}\Lambda$ with $\delta\Pi_{\mu}=0$. Additionally, the mechanism demonstrated in this section can be straightforwardly extended to a more general theory coupled with matter fields, as we will discuss in the next section.

%%%%%%%%%%%%%%%%%%%%%%%%%%%%%%%%%%%%%%%%%%%%%%%%%%%%%%%%%%%%%%%%%%%%%%%%%%%%%%%%
\section{Coupling with fermionic matter}\label{s3}

\subsection{Gauge sector}

Once we have investigated the duality between the free MSC and SD models in the VSR context, the next step is to consider coupling with matter fields. Considering the coupled system of the Dirac field and the above gauge field $A_\mu$, the $SIM(1)$ fermionic Lagrangian density can be expressed as \cite{Alfaro:2019koq,Cheon:2009zx}
\begin{align}\label{fermion}
\mathcal{L}_{\text{fermion}} & =\bar{\psi}\left(i\slashed{D}+i\frac{m_{\psi}^{2}}{2}\frac{\slashed{n}}{n\cdot D}-m\right)\psi,
\end{align}
where $D_{\mu}=\partial_{\mu}-ieA_{\mu}$ is the usual covariant derivative, and $m_{\psi}$ is the VSR-mass associated to the fermionic field.

In addition, due to the presence of the nonlocal operator $1/n\cdot D$, we have a new type of matter current that incorporates nonlinear couplings explicitly dependent on the gauge field. By employing the following matrix identity: 
\begin{equation}
\frac{1}{A+B}=\frac{1}{A}-\frac{1}{A}B\frac{1}{A+B}=\frac{1}{A}-\frac{1}{A}B\frac{1}{A}+\frac{1}{A}B\frac{1}{A}B\frac{1}{A+B}=\cdots,
\end{equation}we can expand this term in the fermionic Lagrangian (\ref{fermion}), resulting in
\begin{align}\label{Lfermi}
\mathcal{L}_{\text{fermion}} & =\bar{\psi}\left(i\slashed{D}+i\frac{m_{\psi}^{2}}{2}\frac{\slashed{n}}{n\cdot D}-m\right)\psi\nonumber\\
 & =\bar{\psi}\left[\left(i\slashed{\partial}+i\frac{m_{\psi}^{2}}{2}\frac{\slashed{n}}{n\cdot\partial}-m\right)\right.\nonumber\\
 & +e\left(\slashed{A}-\frac{m_{\psi}^{2}}{2}\frac{\slashed{n}}{n\cdot\partial}(n\cdot A)\frac{1}{n\cdot\partial}\right)\\ 
 & -ie^{2}\left(\frac{m_{\psi}^{2}}{2}\frac{\slashed{n}}{n\cdot\partial}(n\cdot A)\frac{1}{n\cdot\partial}(n\cdot A)\frac{1}{n\cdot\partial}\right)\nonumber\\
 & \left.+e^{3}\left(\frac{m_{\psi}^{2}}{2}\frac{\slashed{n}}{n\cdot\partial}(n\cdot A)\frac{1}{n\cdot\partial}(n\cdot A)\frac{1}{n\cdot\partial}(n\cdot A)\frac{1}{n\cdot\partial}\right)+\cdots\right]\psi.\nonumber
\end{align}

At this stage, the expression (\ref{Lfermi}) gives rise to an infinite number of vertices with external gauge legs equal to the power in the coupling constant $e$. In what follows, let us restrict ourselves to the case of linear couplings involving fermionic matter and external fields. The cases involving higher powers of the coupling constant are beyond our present scope.

To achieve this, we now define the VSR-SD model linearly coupled to the fermionic field in the following form:
\begin{align}
\mathcal{L}_{\psi-SD}=\mathcal{L}_{SD}+e A_\mu (J^\mu + \mathcal{J}^\mu) + \mathcal{L}(\psi),
\end{align}
where $\mathcal{L}(\psi)$ is the VSR Lagrangian for the free Dirac given by 
\begin{align}
\mathcal{L}(\psi)=\overline{\psi}\left(i\gamma^\mu\partial_\mu+\frac{im_{\psi}^2}{2}\frac{\gamma^\mu n_\mu}{(n\cdot\partial)}-m\right)\psi,   
\end{align}
and the matter currents are denoted by $J^\mu$ and $\mathcal{J}^\mu$, written as
\begin{align}
J^\mu=\overline{\psi}\gamma^\mu\psi,\\
\mathcal{J}^\mu=\frac{1}{2}m_{\psi}^2 n^\mu\bigg(\frac{1}{(n\cdot\partial)}\overline{\psi}\bigg)\slashed{n}\bigg(\frac{1}{(n\cdot\partial)}\psi\bigg).
\end{align}
The first current is the usual one, whereas the second one displays the modification that arises from VSR. To investigate the duality with fermionic matter, let us consider the master Lagrangian (\ref{mlf1}) with linear coupling to the self-dual field:
\begin{align}\label{mlm}
\mathcal{L}_{M}=\chi\theta\epsilon^{\mu\nu\rho}\Pi_\mu\partial_\nu A_\rho+\frac{\chi\theta m_{A}^2}{2}\epsilon^{\mu\nu\rho}\Pi_\mu\frac{n_\rho}{(n\cdot\partial)^2}\partial_\nu (n\cdot A) +\frac{\chi\theta m_{A}^2}{2}\epsilon^{\mu\nu\rho}\Pi_\mu \frac{n_\nu}{(n\cdot \partial)}A_\rho+\frac{\theta^2}{2}\Pi_\mu \Pi^\mu\nonumber\\+\frac{\theta}{2}\epsilon^{\mu\nu\rho}A_\mu\partial_\nu A_\rho-\frac{\theta m_{A}^2}{2}\epsilon^{\mu\nu\rho}A_\mu \frac{n_\nu}{(n\cdot \partial)^2}\partial_\rho(n\cdot A)+\frac{\theta m_{A}^2}{4}\epsilon^{\mu\nu\rho}A_\mu\frac{n_\nu}{(n\cdot \partial)}A_\rho \nonumber\\+e \Pi_\mu (J^\mu + \mathcal{J}^\mu)+\mathcal{L}(\psi). 
\end{align}

By varying the action associated with $\mathcal{L}_M$ with respect to $\Pi_\mu$ and $A_\mu$, we obtain the following equations of motion:
\begin{align}
A_\mu+\frac{m_{A}^2}{2}\frac{n_\mu}{(n\cdot\partial)^2}(n\cdot A)=\Pi_\mu+\partial_\mu \Sigma +\frac{m_{A}^2}{2}\frac{n_\mu}{(n\cdot \partial)} \Sigma \label{am}, \\
\Pi^\mu=\frac{1}{\theta}\epsilon^{\mu\nu\rho}\partial_\nu A_\rho+\frac{m_{A}^2}{2\theta}\epsilon^{\mu\nu\rho}\frac{n_\rho}{(n\cdot \partial)^2}\partial_\nu (n\cdot A)+\frac{m_{A}^2}{2\theta}\epsilon^{\mu\nu\rho}\frac{n_\nu}{(n\cdot \partial)} A_\rho+\frac{e}{\theta^2}(J^\mu + \mathcal{J}^\mu) \label{ppm} .
\end{align}
Substituting (\ref{am}) into (\ref{mlm}), we get the relation
\begin{align}\label{sdf}
\mathcal{L}_{\psi-SD}=\mathcal{L}_{SD}+e \Pi_\mu (J^\mu + \mathcal{J}^\mu)+\mathcal{L}(\psi).
\end{align}
Additionally, inserting (\ref{ppm}) into (\ref{mlm}), we achieve the result
\begin{align}\label{mcsf}
\mathcal{L}_{\psi-MCS}=\mathcal{L}_{MCS}+e\bigg(\frac{1}{\theta}\epsilon^{\mu\nu\rho}\partial_\nu A_\rho+\frac{m_{A}^2}{2\theta}\epsilon^{\mu\nu\rho}\frac{n_\rho}{(n\cdot \partial)}\partial_\nu (n\cdot A)+\frac{m_{A}^2}{2\theta}\epsilon^{\mu\nu\rho}\frac{n_\nu}{(n\cdot \partial)} A_\rho\bigg)(J^\mu + \mathcal{J}^\mu)\nonumber\\-\frac{e^2}{2\theta^2}(J^\mu + \mathcal{J}^\mu)^2+\mathcal{L}(\psi). 
\end{align}

From the above result, we can see that the MCS model acquires a Thirring-like term. This term is modified due to VSR and involves only the matter field. On the other hand, the relation between the models can be expressed through the identification:
\begin{align}\label{pif}
\Pi^\mu=F^\mu-\frac{m_{A}^2}{2}\frac{n^\mu n_\lambda}{(n\cdot\partial)^2}F^\lambda+\frac{e}{\theta^2}(J^\mu + \mathcal{J}^\mu).    
\end{align}
Contrary to the free field case, the identification between $\Pi^{\mu}$ and $F^{\mu}$ fields gains a current term, which will play an important role as we analyze the fermion sector of both models.

%%%%%%%%%%%%%%%%%%%%%%%%%%%%%%%%%%%%%%%
\subsection{Matter sector}

As shown previously, the duality between the MCS and SD models was established by also considering a fermionic matter source. However, we have so far only dealt with the gauge sector. The next step is to verify the duality in the fermion sector. Now, by performing a functional variation on (\ref{sdf}), we obtain:
\begin{align}\label{f1}
\frac{\delta}{\delta\psi}\int d^3x \mathcal{L}_{\psi-SD}=0\Rightarrow\frac{\delta \mathcal{L}(\psi)}{\delta\psi}=-e\Pi_\mu\frac{\delta}{\delta\psi}(J^\mu + \mathcal{J}^\mu).
\end{align}
On the other hand, the gauge sector for SD model with matter source reads 
\begin{align}
\bigg(\theta\eta^{\mu\rho}- \epsilon^{\mu\nu\rho}\partial_{\nu}-\frac{1}{2} m_{f}^2\epsilon^{\mu\nu\rho}\frac{n_\nu}{(n\cdot\partial)}\bigg)\Pi_\rho=-e(J^\mu + \mathcal{J}^\mu).
\end{align}

To express Equation (\ref{f1}) solely in terms of current terms, we must invert the above equation. To achieve this, it is necessary to construct a close algebra.
\begin{align}
\frac{\delta \mathcal{L}(\psi)}{\delta\psi}=\frac{e^2}{\square+m_{A}^2+\theta^2}\bigg(\eta_{\rho\mu}+\frac{\partial_\rho\partial_\mu}{\theta^2}+\frac{m_{A}^2}{2\theta^2}\frac{n_\rho\partial_\mu+n_\mu\partial_\rho}{(n\cdot\partial)}\nonumber\\+\frac{m_{A}^4}{4\theta^2}\frac{n_\rho n_\mu}{(n\cdot\partial)^2}+\epsilon_{\rho\mu\lambda}\frac{\partial^\lambda}{\theta}+\frac{m_{A}^2}{2\theta}\epsilon_{\rho\mu\lambda}\frac{n^\lambda}{(n\cdot\partial)}\bigg)(J^\rho + \mathcal{J}^\rho)\frac{\delta}{\delta\psi}(J^\mu + \mathcal{J}^\mu).    
\end{align}
Hence, we have derived a nonlocal differential equation expressed solely in terms of the matter fields. The result obtained in Ref. \cite{Gomes:1997mf} can be recovered in the case of $m_{A}=0$. Now, let us shift our focus to the MCS model by varying (\ref{mcsf}):
\begin{align}\label{efm1}
\frac{\delta}{\delta\psi}\int d^3x \mathcal{L}_{\psi-MCS}=0\Rightarrow\frac{\delta \mathcal{L}(\psi)}{\delta\psi}=-\frac{e^2}{\theta^2}(J^\mu + \mathcal{J}^\mu)\frac{\delta}{\delta\psi}(J^\mu + \mathcal{J}^\mu)\nonumber\\-e\bigg(F^\mu-\frac{m_{A}^2}{2}\frac{n^\mu n_\lambda}{(n\cdot\partial)^2}F^\lambda\bigg)\frac{\delta}{\delta\psi}(J^\mu + \mathcal{J}^\mu).
\end{align}
Inserting (\ref{pif}) into above equation, we get
\begin{align}
\frac{\delta \mathcal{L}(\psi)}{\delta\psi}=-e\Pi_\mu\frac{\delta}{\delta\psi}(J^\mu + \mathcal{J}^\mu),
\end{align}
which can be rewritten as
\begin{align}\label{efm2}
\frac{\delta \mathcal{L}(\psi)}{\delta\psi}=\frac{e^2}{\square+m_{A}^2+\theta^2}\bigg(\eta_{\rho\mu}+\frac{\partial_\rho\partial_\mu}{\theta^2}+\frac{m_{A}^2}{2\theta^2}\frac{n_\rho\partial_\mu+n_\mu\partial_\rho}{(n\cdot\partial)}\nonumber\\+\frac{m_{A}^4}{4\theta^2}\frac{n_\rho n_\mu}{(n\cdot\partial)^2}+\epsilon_{\rho\mu\lambda}\frac{\partial^\lambda}{\theta}+\frac{m_{A}^2}{2\theta}\epsilon_{\rho\mu\lambda}\frac{n^\lambda}{(n\cdot\partial)}\bigg)(J^\rho + \mathcal{J}^\rho)\frac{\delta}{\delta\psi}(J^\mu + \mathcal{J}^\mu).    
\end{align}

From the above results, we can see that the fermionic sector described by SD in Eq. (\ref{efm1}) is equivalent to the fermionic sector related to MCS in Eq. (\ref{efm2}). Consequently, we have thoroughly established the classical duality between MCS and SD coupled to fermionic matter in the VSR context. As a final statement,  it is worth pointing out that the result found in the literature regarding the duality with the fermionic sector is recovered when turning off the VSR parameters.

%%%%%%%%%%%%%%%%%%%%%%%%%%%%%%%%%%%%%%%%%%%%%%%%%%%%%%%%%%%%%%%%%%%%
%\section{Duality at quantum level}
%Now that the duality between Maxwell-Chern-Simons and Self-Dual models has been proved at classical level, we can proceed to investigate the duality at quantum level.

%\begin{align}
%Z=\int\mathcal{D}A\mathcal{D}\Pi e^{i\int d^3x \mathcal{L}}   
%\end{align}

%%%%%%%%%%%%%%%%%%%%%%%%%%%%%%%%%%%%%%%%%%%%%%%%%%%%%%%%%%%%%%%%%%%%%%%%%%%%%%%%%%%%%%%%%%%
\section{Conclusion}\label{s4}

The main goal of this work was to establish the duality between the Maxwell-Chern-Simons (MCS) model and the first-order derivative theory known as self-dual within the framework of very special relativity. Initially, we constructed both models by deriving the equations of motion and propagators. Subsequently, we demonstrated equivalence at the level of equations of motion, focusing solely on the free-field case. In the sequel, we employed an approach known as the master action, allowing us to construct a Lagrangian density capable of connecting both MCS and self-dual (SD) theories. This method provided a direct proof of equivalence at the classical level in the VSR scenario. Furthermore, we extended our analysis to include the matter field with a fermion source. In the presence of interaction, we show that the MCS model acquires a Thirring-like interaction corrected by non-local VSR terms to maintain the established duality.

%%%%%%%%%%%%%%%%%%%%%%%%%%%%%%%%%%%%%%%%%%%%%%%%%%%%%%%%%%%%%%%%%%%%%%%%%%%%%%%%%%%%%%%%%%%

%%%%%%%%%%%%%%%%%%%%%%%%%%%%%%%%%%%%%%%%%%%%%%%%%%%%%%%%%%%%%%%%%%%%%%%%%%
\section*{Acknowledgments}
\hspace{0.5cm} The authors thank the Funda\c{c}\~{a}o Cearense de Apoio ao Desenvolvimento Cient\'{i}fico e Tecnol\'{o}gico (FUNCAP), the Coordena\c{c}\~{a}o de Aperfei\c{c}oamento de Pessoal de N\'{i}vel Superior (CAPES), and the Conselho Nacional de Desenvolvimento Cient\'{i}fico e Tecnol\'{o}gico (CNPq), Grant no. 200879/2022-7 (RVM). R. V. Maluf thanks the Department of Theoretical Physics \& IFIC  of the University of Valencia - CSIC for the kind hospitality.

%%%%%%%%%%%%%%%%%%%%%%%%%%%%%%%%%%%%%%%%%%%%%%%%%%%%%%%%%%%%%%%%%%%%%%%%%%

%%%%%%%%%%%%%%%%%%%%%%%%%%%%%%%%%%%%%%%%%%%%%%%%%%%%%%%%%%%%%%%%%%%%%%%%%%%%

\end{document}